\documentclass[aps,pre,twocolumn,showpacs]{revtex4}
\usepackage{graphicx}
\usepackage{amsmath,amscd,amssymb}
\usepackage{color}

\newcommand{\be}{\begin{equation}}
\newcommand{\ee}{\end{equation}}
\newcommand{\bea}{\begin{eqnarray}}
\newcommand{\eea}{\end{eqnarray}}
\newcommand{\lan}{\left\langle}
\newcommand{\ran}{\right\rangle}
\newcommand{\br}{\mathbf{r}}
\newcommand{\ba}{\mathbf{a}}

\newcommand{\bb}{\mathbf{b}}
\newcommand{\bq}{\mathbf{q}}

\newcommand{\bu}{\mathbf{v}}

\newcommand{\bx}{\mathbf{x}}
\newcommand{\by}{\mathbf{y}}
\newcommand{\bo}{\mathbf{\Omega}_k}
\newcommand{\bom}{\mathbf{\Omega}}

\newcommand{\e}{\varepsilon}

\newcommand{\tv}{\tilde{v}}
\newcommand{\te}{\tilde{\epsilon}}

\begin{document}

\title{Alteration of gas phase ion polarizabilities upon hydration in high dielectric liquids}

\author{Sahin Buyukdagli$^{1}$\footnote{email:~\texttt{sahin\_buyukdagli@yahoo.fr}} and T. Ala-Nissila$^{1,2}$\footnote{email:~\texttt{Tapio.Ala-Nissila@aalto.fi}}}
\affiliation{$^{1}$Department of Applied Physics and COMP Center of Excellence, Aalto University School of Science, P.O. Box 11000, FI-00076 Aalto, Espoo, Finland\\
$^{2}$Department of Physics, Brown University, Providence, Box 1843, RI 02912-1843, U.S.A.}
\date{\today}

\begin{abstract}
We investigate the modification of gas phase ion polarizabilities upon solvation in polar solvents and ionic liquids. To this aim, we develop a classical electrostatic theory of charged liquids composed of solvent molecules modeled as finite size dipoles, and embedding polarizable ions that consist of Drude oscillators. In qualitative agreement with ab-initio calculations of polar solvents and ionic liquids, the hydration energy of a polarizable ion in both type of dielectric liquid is shown to favor the expansion of its electronic cloud. Namely, the ion carrying no dipole moment in the gas phase acquires a dipole moment in the liquid environment, but its electron cloud also reaches an enhanced rigidity. We find that the overall effect is an increase of the gas phase polarizability upon hydration. In the specific case of ionic liquids, it is shown that this hydration process is driven by a collective solvation mechanism where the dipole moment of a polarizable ion induced by its interaction with surrounding ions self-consistently adds to the polarization of the liquid, thereby amplifying the dielectric permittivity of the medium in a substantial way.  We propose this self-consistent hydration as the underlying mechanism behind the high dielectric permittivities of ionic liquids composed of small charges with negligible gas phase dipole moment. Hydration being a correlation effect, the emerging picture indicates that electrostatic correlations cannot be neglected in polarizable liquids.
\end{abstract}

\maketitle
\section{Introduction}
The atomic electron cloud distortion induced by an external field is strongly influenced by the dielectric environment embedding the atom. This distortion ability referred as the \textit{induced polarizability} is one of the key ion specific effects in the simulation of salt solutions in inhomogeneous media such as the water-air interface or protein-water surfaces~\cite{jun}. The precise knowledge of the change in the polarizability of an isolated ion upon hydration in water is particularly important for the development of polarizable force fields used in these simulations. Moreover, ionic polarizability is also believed to have a substantial effect on the polarity of ionic liquids. Indeed, numerical studies based on ab-initio calculations show that the large dielectric permittivity of ionic liquids such as $[\mathrm{C}_2\mathrm{mim}][\mathrm{NTf}_2]$ and $[\mathrm{C}_2\mathrm{mim}][\mathrm{BF}_4]$ composed of ions with small individual dipole moments cannot be solely explained by their rotational polarizability~\cite{ionion2}. This suggests that an additional polarization mechanism resulting from the interaction of the polarizable ion with the surrounding ions in the liquid must be present.

The alteration of ionic gas phase polarization upon solvation has been so far considered  within numerical approaches based on quantum calculations with polarizable continuum model (PCM) or explicit solvent. These two approaches interestingly yield diverging pictures on the hydration of polarizable ions. Namely, the calculations with explicit solvent indicate that the ionic polarizability is decreased with respect to the gas phase~\cite{jun2}, whereas PCM approaches yield a higher polarizability in the liquid state~\cite{ionpol,ionpol2} (see also Ref.~\cite{NetzRev} for a review on the computational state of the art). The latter case is also in line with the ab-initio calculations of pure water clusters~\cite{watwat} and ionic liquids~\cite{ionion1}, where the transfer of both type of molecules from gas to the liquid environment was shown to increase their dipole moment.

In order to understand the physics behind the hydration of polarizable molecules, analytical theories offering a deeper understanding are needed. The theoretical formulation of the problem requires in turn an explicit and realistic consideration of the discrete charge structure of solvent molecules and ions. Unfortunately, this level of refinement has been until recently beyond the state of the art of electrostatic theories, which are mostly based on dielectric continuum solvents embedding point charges. The first statistical theory of inhomogeneous electrolytes with explicit solvent was introduced in Ref.~\cite{dundip} in the form of a mean-field (MF) dipolar Poisson-Boltzmann (DPB) equation. This approach that models the solvent molecules as point dipoles was later generalized by including the steric interactions between the particles for inhomogeneous charged liquids~\cite{orland1}, and a one-loop extension was presented as well in Ref.~\cite{orland2} to explain the salt induced dielectric decrement effect in bulk electrolytes. We have recently incorporated into the DPB approach surface polarization effects, which allowed us to significantly improve the agreement of the dielectric continuum electrostatic with experimental capacitance data of carbon based materials~\cite{epl}. Sophisticated electrostatic formulations accounting for the dipolar and higher order multipolar moments of ions in the point dipole limit have been also proposed in Refs.~\cite{netzvdw,bohinc1,podgornik,dem}. In a similar context, we can also mention the works of Refs.~\cite{Lue1,pincus,bohinc2} where the extended charge structure of rigid linear molecules was ingeniously considered.

We have recently developed a non-local electrostatic theory of polar liquids with explicit solvent and polarizable ions beyond the point dipole approximation~\cite{nlpb}. The electrolyte model that treats solvent molecules as finite size dipoles and polarizable ions as Drude oscillators was investigated at the MF level. It was shown that the consideration of the extended charge structure of solvent molecules enables us to capture the non-local dielectric response of water at charged interfaces observed in molecular dynamics simulations and atomic force experiments. In this article, we reconsider the model of Ref.~\cite{nlpb} beyond the MF level of approximation in order to characterize the hydration induced modification of ionic polarizabilities in high dielectric bulk liquids. We review in Section~\ref{mod} the derivation of the field theoretic charged liquid model, and derive the closure equations accounting for the correlations between the ions and the solvent molecules. These equations are first solved in Section~\ref{polar} in order to investigate the hydration of a single polarizable ion in a polar solvent such as water. Then, within the same theoretical framework, we consider in Section~\ref{ion} an ionic liquid free of solvent molecules in order to investigate a collective polarization effect in the liquid. It is shown that in both systems, our simple theory can capture the solvation induced electronic cloud expansion effect observed in ab-initio calculations~\cite{ionpol,ionpol2,ionion1}, and provides a physical explanation in terms of the electrostatic energy released by the ion upon hydration. The limitations of the liquid model and the computation scheme, and necessary extensions are discussed in detail in the Conclusion.

\section{Model}
\label{mod}

We briefly review in this section the derivation of the field theoretic partition function for the polar liquid model previously introduced in Ref.~\cite{nlpb}. Then, starting from the Dyson equation, we derive an integral equation for the dielectric permittivity function embodying the interactions between the polarizable ions and solvent molecules of the bulk liquid.

\begin{figure}
\includegraphics[width=1.0\linewidth]{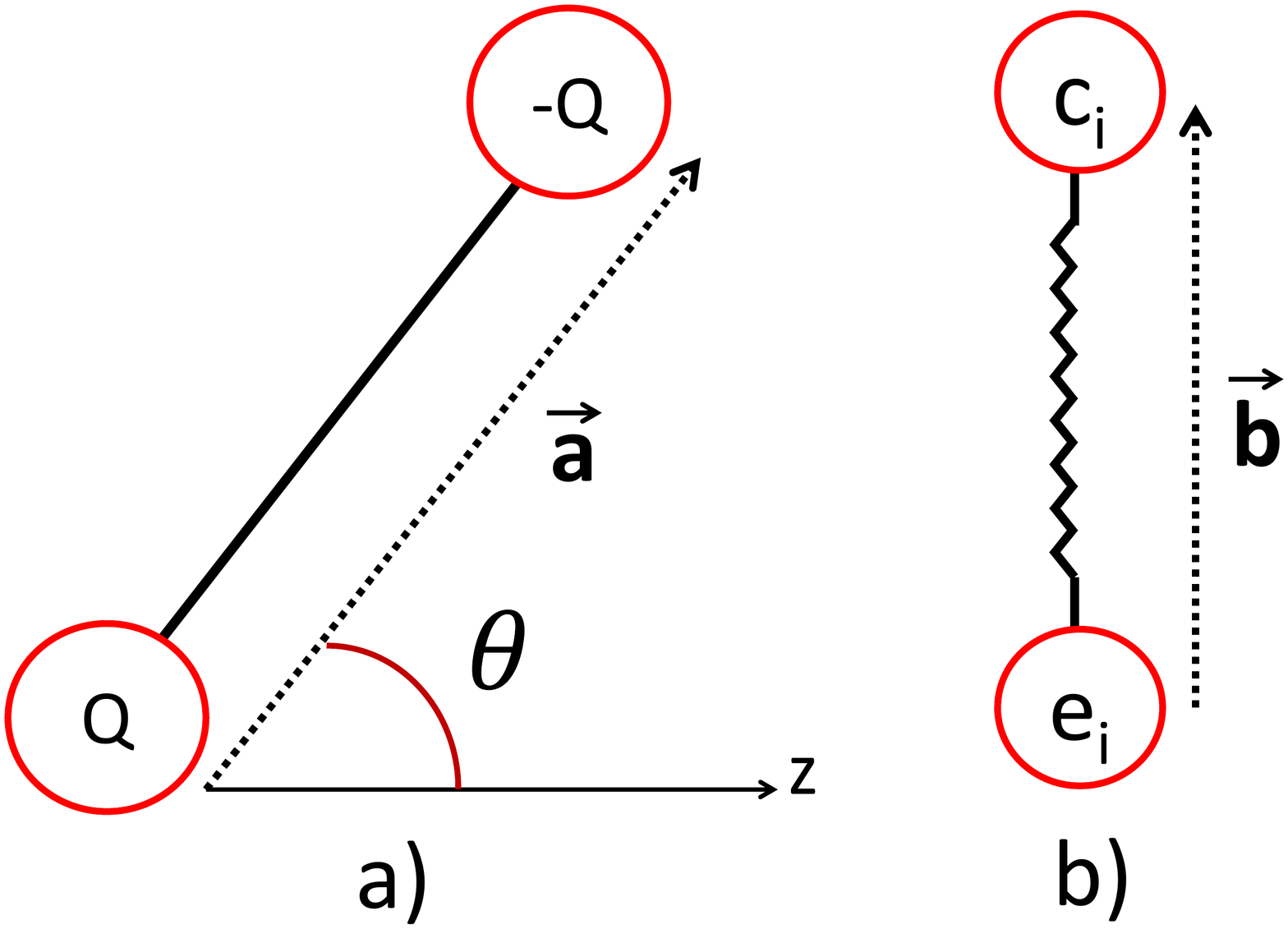}
\caption{(Color online) Charge composition of solvent molecules of size $a$ (a) and polarizable ions with a fluctuating length $b$ (b). In the present work, we consider exclusively the case of ionic valencies $e_i$ and $c_i$ of opposite sign ($e_ic_i<0$), and solvent molecules with monovalent point charges $Q=1$.}
\label{fig0}
\end{figure}

The geometry of solvent molecules is depicted in Fig.~\ref{fig0}(a). The polar liquid is composed of overall neutral solvent molecules modeled as linear dipoles of length $a$, and two point charges of valency $\pm Q=\pm1$ at the extremities. Furthermore, the solvent contains polarizable molecules of $p$ species, each of them being an oscillating rod of length $b$ (see Fig.~\ref{fig0}(b)). The point charges $e_i$ and $c_i$ at the extremities satisfy the inequality $e_ic_i<0$, where the index $i=1...p$ runs over the ionic species. Moreover, the ionic polarizability is taken into account within the Drude oscillator model~\cite{drude},
\be\label{hpol}
h_i\left(\bb\right)=\frac{\bb^2}{4b_{pi}^2},
\ee
where the square of the variance of electronic cloud oscillations $b_{pi}^2$ is proportional to the induced polarizability of ions $\alpha$ in the gas phase~\cite{nlpb}. Because the former offers a more intuitive realization of the electronic cloud fluctuations induced by thermal excitations, we will discuss the results in terms of the length scale $b_{pi}$. Furthermore, in the present work, we will consider exclusively the case of equal ionic polarizabilities for all species, but the analytical results will be given for the general case. We also note that the electroneutrality condition implies the equality $\sum_i\rho_{ib}q_i=0$, with $\rho_{ib}$ the bulk density, and $q_i=e_i+c_i$ the total charge of the polarizable molecules with species $i$.

The canonical partition function for the system composed of solvent molecules and ions coupled with electrostatic interactions read
\bea\label{canpart}
Z_c&=&\frac{e^{N_s\epsilon_s}}{N_s!\lambda_{Td}^{3N_s}}\int\prod_{k=1}^{N_s}\frac{\mathrm{d}\bo}{4\pi}\mathrm{d}\bx_k\\
&&\times\prod_{i=1}^{p}\prod_{j=1}^{N_i}\frac{e^{N_i\epsilon_i}}{N_i!\lambda_{Ti}^{3N_i}}\int\frac{\mathrm{d}\bb_j}{\left(4\pi b_{pi}^2\right)^{3/2}}\mathrm{d}\by_{ij}\;e^{-h_i\left(\bb_j\right)-H(\bu)},\nonumber
\eea
where $N_s$ is the total number of solvent molecules, $N_i$ the number of ions for the species $i$, and $\lambda_{Td}$ and $\lambda_{Ti}$ denote respectively the thermal wavelengths of solvent molecules and ions. We also introduced in Eq.~(\ref{canpart}) the compact notation $\bu=\left(\{\bx_k\},\{\ba_k\},\{\by_{ij}\},\{\bb_j\}\right)$ for the vectors characterizing the configuration of particles, with $\bx_k$ and $\by_{ij}$ denoting respectively the coordinate of the charges $+Q$ and $e_i$ of the solvent molecules and polarizable ions in depicted in Fig.~\ref{fig0}. Furthermore, $\bo=(\theta_k,\varphi_k)$ is the solid angle characterizing the orientation of the $kth$ solvent molecule, $\theta$ being the angle between the oriented dipole and the $z$-axis (see Fig.~\ref{fig0}(a)). We finally note that in Eq.~(\ref{canpart}), we subtracted from the total Hamiltonian the self energies of ions and polar molecules in the air, $\epsilon_i=\left(e_i^2+c_i^2\right)v_c(0)/2+e_ic_iv_c(b)$ and $\epsilon_s=Q^2\left[v_c(0)-v_c(a)\right]$. This point will be discussed below in further detail.

The Hamiltonian of the bulk liquid is composed of pairwise electrostatic interactions,
\be\label{Hel}
H_{el}(\bu)=\frac{1}{2}\int_{\br\br'}\left[\rho_{ic}+\rho_{sc}\right]_\br v_c(\br-\br')\left[\rho_{ic}+\rho_{sc}\right]_{\br'},
\ee
where the total ionic and solvent density operators for the charge compositions depicted in Fig.~\ref{fig0} are defined as
\bea\label{ci}
&&\rho_{ic}(\br)=\sum_{i=1}^p\sum_{j=1}^{N_i}\left[e_i\delta(\br-\by_{ij})+c_i\delta(\br-\by_{ij}-\bb_j)\right]\\
\label{cd}
&&\rho_{sc}(\br)=\sum_{k=1}^{N_s}Q\left[\delta(\br-\bx_k)-\delta(\br-\bx_k-\ba_k)\right].
\eea
Moreover, in Eq.~(\ref{Hel}), $v_c(\br-\br')=\ell_B/|\br-\br'|$ stands for the Coulomb potential in the air medium, with $\ell_B=e^2/\left[4\pi\e_0k_BT\right]\simeq 55$ nm the Bjerrum length and $\e_0$ the dielectric permittivity in the air, $e$ the electron charge, and $T=300$ K the ambient temperature. We note that in the rest of the article, dielectric permittivities will be expressed in units of the air permittivity $\e_0$, and energies in units of the thermal energy $k_BT$.

In order to transform the partition function~(\ref{canpart}) into a more tractable form, we pass from the particle density to the fluctuation potential representation by performing a standard Hubbard-Stratonovich transformation. In this representation, the grand canonical partition function of the system defined as $Z_G=\sum_{N_s\geq0}\prod_{i=1}^p\sum_{N_i\geq0}e^{\mu_i N_i}e^{\mu_w N_s}Z_c$ takes the form of a functional integral over the fluctuating electrostatic potential $\phi(\br)$, $Z_G=\int \mathcal{D}\phi\;e^{-H[\phi]}$, with the Hamiltonian functional~\cite{nlpb}
\bea\label{HamFunc}
H[\phi]&=&\int\mathrm{d}\br\frac{\left[\nabla\phi(\br)\right]^2}{8\pi\ell_B}-\Lambda_s\int\frac{\mathrm{d}\bom}{4\pi}\mathrm{d}\br\;e^{\epsilon_s+iQ\left[\phi(\br)-\phi(\br+\ba)\right]}\nonumber\\
&&-\sum_i\Lambda_i\int\frac{\mathrm{d\bb}}{\left(4\pi b_{pi}^2\right)^{3/2}}\mathrm{d}\br\;e^{-h_i(\bb)+\epsilon_i}\nonumber\\
&&\hspace{3.5cm}\times e^{ie_i\phi(\br)+ic_i\phi(\br+\bb)}.
\eea
The first term on the r.h.s. of Eq.~(\ref{HamFunc}) is the electrostatic energy of the freely propagating field in the air. The second term corresponds to the density of solvent molecules, and their fugacity is denoted by $\Lambda_s$. Finally, the third term on the r.h.s. of Eq.~(\ref{HamFunc}) is the density of polarizable ions with fugacity $\Lambda_i$.

The Hamiltonian~(\ref{HamFunc}) was already derived in Ref.~\cite{nlpb} for the more general case of multipolar solvents embedding polarizable ions, and the saddle-point solution of the partition function corresponding to the MF approximation was investigated for polar liquids in contact with charged planes. In order to account for correlation effects in the bulk liquid beyond the MF level, we need to derive the electrostatic correlation function. Our starting point is the following form of the Dyson equation,
\be\label{Dyson}
\int\mathcal{D}\phi\frac{\delta}{\delta\phi(\br)}\;e^{-H[\phi]+\int\mathrm{d}\br J(\br)\phi(\br)}=0,
\ee
where $J(\br)$ is a generalized current introduced for the derivation of the two point correlation function. A proof of the equality~(\ref{Dyson}) can be found in Ref.~\cite{justin}. We also remind that the derivation of the electrostatic self-consistent equations of the primitive ion model~\cite{netzvar} with the use of this equality was presented in Ref.~\cite{jcp}. By taking now the functional derivative of Eq.~(\ref{Dyson}) with respect to $J(\br')$ and setting $J(\br')=0$, one obtains the following equation for the two point correlation function,
\bea\label{corr2}
&&\nabla_\br^2\lan\phi(\br)\phi(\br')\ran\\
&&+4\pi\ell_BiQ\lambda_s\int\frac{\mathrm{d}\bom}{4\pi}\mathrm{d}\br\;e^{\epsilon_s}\;\left\{\lan e^{iQ\left[\phi(\br)-\phi(\br+\ba)\right]}\phi(\br')\ran\right.\nonumber\\
&&\hspace{4cm}-\left.\lan e^{iQ\left[\phi(\br-\ba)-\phi(\br)\right]}\phi(\br')\ran\right\}\nonumber\\
&&+4\pi\ell_Bi\sum_i\Lambda_i\int\frac{\mathrm{d\bb}}{\left(4\pi b_{pi}^2\right)^{3/2}}\mathrm{d}\br\;e^{-h_i(\bb)+\epsilon_i}\nonumber\\
&&\times\left\{\lan\left[e_ie^{ie_i\phi(\br)+ic_i\phi(\br+\bb)}+c_ie^{ie_i\phi(\br-\bb)+ic_i\phi(\br)}\right]\phi(\br')\ran\right\}\nonumber\\
&&=-4\pi\ell_B\delta(\br-\br')\nonumber,
\eea
where the bracket $\lan\cdot\ran$ denotes the field average with the Hamiltonian Functional in Eq.~(\ref{HamFunc}). In Eq.~(\ref{corr2}), the dependence of the fluctuating solvent and ion densities (i.e. the functions inside the brackets on the l.h.s.) on the values of the potential at different points around $\br$ is a signature of non-local electrostatic interactions resulting from the extended charge structure of the solvent molecules and ions~\cite{nlpb}.

We emphasize that the formal equation~(\ref{corr2}) is an exact relation. However, because the Hamiltonian of Eq.~(\ref{HamFunc}) is non-linear in the potential $\phi(\br)$, an exact analytical evaluation of the averages over the fluctuating potential is impossible. To progress further, we approximate this non-linear Hamiltonian with a quadratic Hamiltonian functional,
\be\label{gauss}
H_0[\phi]=\int\frac{\mathrm{d}\br\mathrm{d}\br'}{2}\phi(\br)v_0^{-1}(\br,\br')\phi(\br),
\ee
where the electrostatic potential is chosen as the solution of the equation~(\ref{corr2}), that is, $v_0(\br,\br')=\lan\phi(\br)\phi(\br')\ran$. At this stage, we note that the spherical symmetry in the bulk liquid implies $v_0(\br,\br')=v_0(\br-\br')$, and this allows us to expand the potential in Fourier space as $v_0(\br-\br')=\int\frac{\mathrm{d}^3\bq}{\left(2\pi\right)^3}\;e^{i\bq\cdot(\br-\br')}\tv_0(q)$. Evaluating the averages in Eq.~(\ref{corr2}) with the quadratic functional~(\ref{gauss}) and injecting into the result the Fourier expansion of the correlation function, the explicit form of the potential finally follows in the form~\cite{rem}
\be\label{varpot}
\tv_0^{-1}(q)=\frac{q^2\te(q)}{4\pi\ell_B}+\sum_i\rho_{ib}q_i^2,
\ee
with the Fourier transformed dielectric permittivity function
\bea\label{varep}
\te(q)&=&1+\frac{\kappa_s^2}{q^2}\left[1-\frac{\sin(qa)}{qa}\right]+\sum_i\frac{\kappa_{ip}^2}{q^2}\lan1-\frac{\sin(qb)}{qb}\ran.\nonumber\\
\eea
We introduced in Eq.~(\ref{varep}) the solvent and ionic screening parameters in the air, $\kappa^2_s=8\pi Q^2\ell_B\rho_{sb}$ and $\kappa^2_{ip}=8\pi|e_ic_i|\ell_B\rho_{ib}$. Furthermore, we defined in Eq.~(\ref{varep}) the statistical average over the fluctuations of the electronic cloud,
\be\label{av}
\lan F(b)\ran=\frac{\int_0^\infty\mathrm{d}bb^2 \;e^{-h_i(b)-\psi_{ip}(b)}F(b)}{\int_0^\infty\mathrm{d}bb^2\;e^{-h_i(b)-\psi_{ip}(b)}},
\ee
with the potential of mean force (PMF)
\be\label{dippmf}
\psi_{ip}(b)=-|e_ic_i|\int_0^\infty\frac{\mathrm{d}qq^2}{2\pi^2}\left[1-\frac{\sin(qb)}{qb}\right]\left[\tv_c(q)-\tv_0(q)\right],\\
\ee
where the Fourier transform of the Coulomb potential in the vacuum given by $\tv_c(q)=q^2/(4\pi\ell_B)$. We also note that deriving Eq.~(\ref{varpot}), we used the thermodynamic relations between the particle fugacities and concentrations, $\rho_{sb}=\Lambda_s\partial\ln Z/\partial\Lambda_s=\Lambda_se^{-\psi_d}$ and $\rho_{ib}=\Lambda_i\partial\ln Z/\partial\Lambda_i=\Lambda_i\int\mathrm{d}\bb e^{-h_i(\bb)-\psi_{ip}(b)}/(4\pi b_{pi}^2)^{3/2}$, with the liquid state self-energies of solvent molecules and ions respectively defined as
\bea\label{PMFs}
&&\psi_s=-Q^2\int_0^\infty\frac{\mathrm{d}qq^2}{2\pi^2}\left[1-\frac{\sin(qa)}{qa}\right]\left[\tv_c(q)-\tv_0(q)\right]\nonumber\\
&&\\
\label{PMFi}
&&\psi_i(b)=-\int_0^\infty\frac{\mathrm{d}qq^2}{2\pi^2}\left\{\frac{e_i^2+c_i^2}{2}+e_ic_i\frac{\sin(qb)}{qb}\right\}\\
&&\hspace{2.9cm}\times\left[\tv_c(q)-\tv_0(q)\right].\nonumber
\eea
One can notice that the energies in Eq.~(\ref{PMFs}) and~(\ref{PMFi}) correspond to the hydration energies of the solvent molecules and polarizable ions, i.e. the electrostatic cost to drive the molecules from the gas to the liquid environment. Moreover, one sees that Eqs.~(\ref{dippmf}) and~(\ref{PMFi}) are related as  $\psi_i(b)=-q_i^2\left[v_c(0)-v_0(0)\right]/2+\psi_{ib}(b)$, which indicates that the PMF $\psi_{ip}(b)$ brings the net contribution from the polarizability to the ionic hydration energy. Finally, unlike previous point dipole models where the electrostatic energies have to be regularized with an ultraviolet cut-off in Fourier space~\cite{orland2,dem}, our consideration of the finite solvent molecular size and electronic cloud extension resulted in a cut-off free theory with well defined self energies in Eqs.~(\ref{dippmf})-(\ref{PMFi}).

At this stage, we note that our motivation for subtracting from the Hamiltonian the gas phase self-energy of polarizable ions in Eq.~(\ref{canpart}) was twofold. First of all, this step allowed us to avoid the dipolar catastrophy problem. Indeed, the classical Drude oscillator model of Eq.~(\ref{hpol}) does not prevent the electron from falling into the nucleus, and this results in divergent ionic self-energies for $b\to 0$. This problem could be avoided in an alternative way by modifying the Drude model with a cut-off at small $b$, but we found that this technical complication shadows the transparency of the analytical results. Furthermore, the Drude potential is clearly an approximative fashion to consider the quantum mechanical interatomic interactions that already include the electrostatic coupling between the electron and the nucleus. We also note that the subtracted self-energy of solvent molecules does not affect the statistical average in Eq.~(\ref{av}).

The relations~(\ref{varpot})-(\ref{dippmf}) form a set of closure equations that should be solved self-consistently. These two relations can be also interpreted as a single integral equation for the dielectric permittivity function $\te(q)$ in Fourier space. Then, one notes that computing the average in Eq.~(\ref{varep}) by neglecting the PMF~(\ref{dippmf}) in Eq.~(\ref{av}), one obtains the MF permittivity function derived in Ref.~\cite{nlpb},
\bea\label{varepmf}
\te_{MF}(q)&=&1+\frac{\kappa_s^2}{q^2}\left[1-\frac{\sin(qa)}{qa}\right]+\sum_i\frac{\kappa_{ip}^2}{q^2}\left[1-e^{-b_{pi}^2q^2}\right].\nonumber\\
\eea
Hence, electrostatic correlation effects are incorporated in the hydration PMF $\psi_{ib}(b)$. In the rest of the article, the solution of the closure equations~(\ref{varpot})-(\ref{dippmf}) will be considered in order to investigate the solvation of polarizable ions in high dielectric liquids.

\section{Results}

In this section, we solve the closure equations~(\ref{varpot})-(\ref{dippmf}) in order to shed light on the electrostatic mechanism behind the hydration effects observed in ab-initio calculations for polarizable ions in high dielectric liquids such as polar solvents~\cite{ionpol,ionpol2} and ionic liquids~\cite{ionion1,ionion2}. We first investigate in Section~\ref{polar} the hydration of a single polarizable ion in a polar liquid such as water, and we characterize in Section~\ref{ion} a similar cooperative solvation mechanism in ionic liquids exclusively composed of polarizable ions.

\subsection{Hydration of a single polarizable ion in water}
\label{polar}
\begin{figure}
\includegraphics[width=1.0\linewidth]{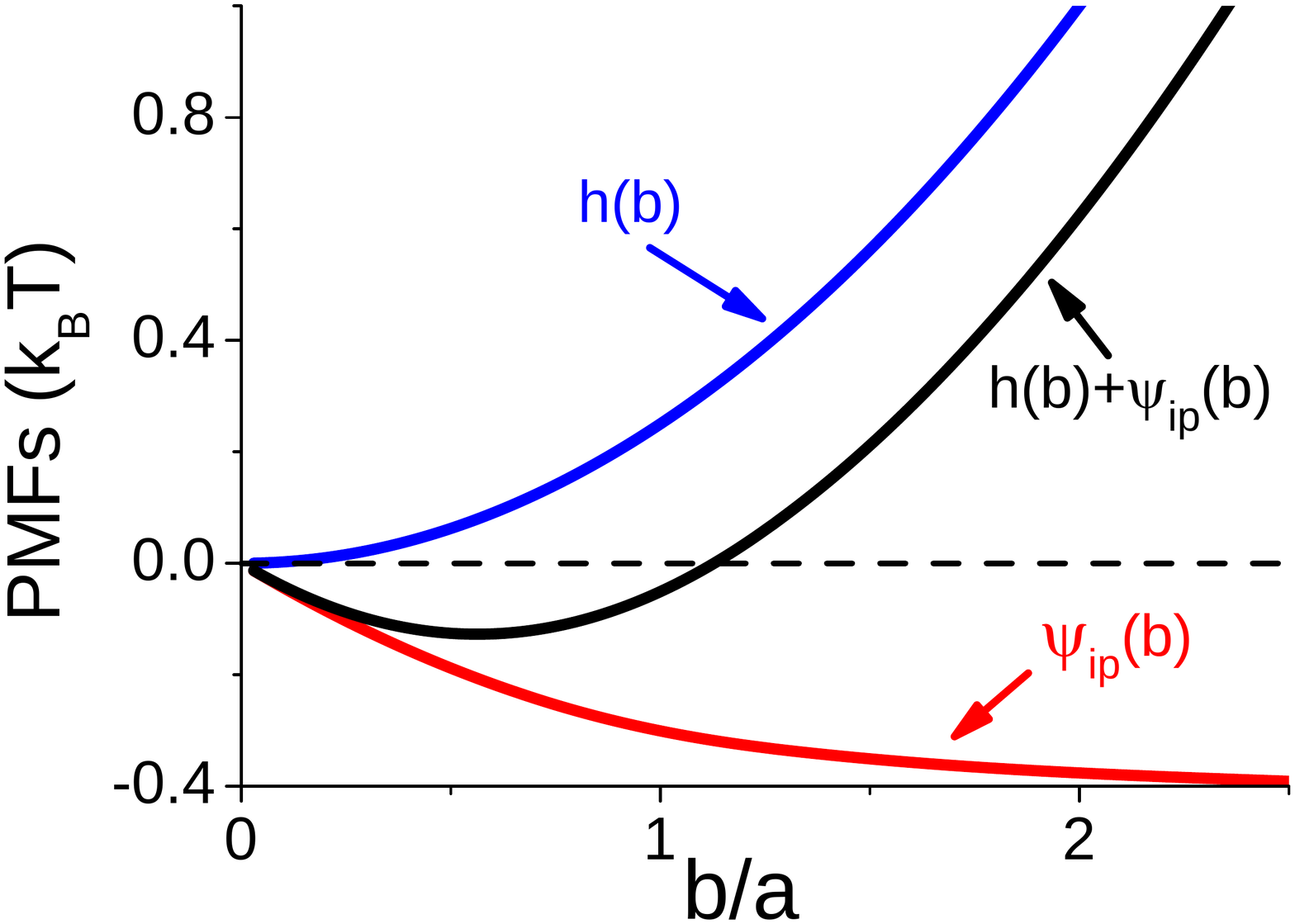}
\caption{(Color online) Drude oscillator potential Eq.~(\ref{hpol}) (blue curve), hydration energy of Eq.~(\ref{pmfdil}) (red curve), and total distortion potential of an hydrated polarizable ion (black curve). Model parameters are $a=1$ {\AA}, $\rho_{sb}=10^{-4}$ M, $b_{pi}=1$ {\AA}, and $|e_ic_i|=2$.}
\label{fig1}
\end{figure}

This section is devoted to the hydration of a single polarizable ion in a strongly polar liquid such as water. In the dilute ion regime, the PMF of Eq.~(\ref{dippmf}) has to be evaluated at the leading order in the ion concentration by neglecting the ionic contributions corresponding respectively to the second and third terms on the r.h.s. of Eqs.~(\ref{varpot}) and~(\ref{varep}). In order to illustrate the hydration mechanism in an intuitive way, we first consider a polarizable ion in a dilute solvent. By expanding Eq.~(\ref{dippmf}) at the order $O\left((\kappa_sa)^2\right)$, which is valid for the solvent molecular size $a=1$ {\AA} in the solvent density regime $\rho_{sb}\lesssim0.1$ M, one obtains for the PMF associated with the polarizability the close form expression,
\bea\label{pmfdil}
\psi_{ip}(b)&=&-|e_ic_i|\frac{\ell_B}{2a}\left(\kappa_sa\right)^2\left\{\frac{3ab-b^2}{3a^2}\theta(a-b)\right.\\
&&\hspace{2.8cm}+\left.\frac{3ab-a^2}{3ab}\theta(b-a)\right\}.\nonumber
\eea

The hydration potential $\psi_{ip}(b)$ of Eq.~(\ref{pmfdil}) and the total distortion energy $h_i(b)+\psi_{ip}(b)$ are compared in Fig.~\ref{fig1} with the distortion potential of an isolated ion $h_i(b)$. One sees that the negative hydration potential $\psi_{ip}(b)$ results in a net reduction of the bare distortion energy $h_i(b)$. In other words, the hydration of a polarizable ion favors the expansion of its electronic cloud. This peculiarity results from the fact that the Born energy of a point charge is proportional to the square of its valency, and the point charges on the polarizable ion are of opposite sign and satisfy the inequality $e_i^2+c_i^2>q_i^2$. As a result, the solvation energy of two separate charges with valencies $e_i$ and $c_i$ is lower than the Born energy of a single ion of valency $q_i$ in Eq.~(\ref{PMFi}), that is $\psi_i(b\to\infty)<\psi_i(b=0)$. It follows from this remark that for a rodlike molecule with the charges $e_i$ and $c_i$ of the same sign, hydration would in turn lead to a compression of the electronic cloud. Furthermore, the black curve in Fig.~\ref{fig1} shows that the total distortion potential $h_i(b)+\psi_{ip}(b)$ exhibits a minimum. This means that the polarizable molecule without average dipole in the gas phase acquires a net dipole moment upon hydration. One finally notes that in Eq.~(\ref{pmfdil}), the hydration potential converges for $b\gtrsim a$ to a constant value $\psi_{ib}=-|e_ic_i|\left(\kappa_sa\right)^2\ell_B/(2a)$.  Thus, for dilute solvents, the hydration modifies the electronic cloud rigidity mainly at separation distances below the solvent molecular size.

\begin{figure*}
\includegraphics[width=.47\linewidth]{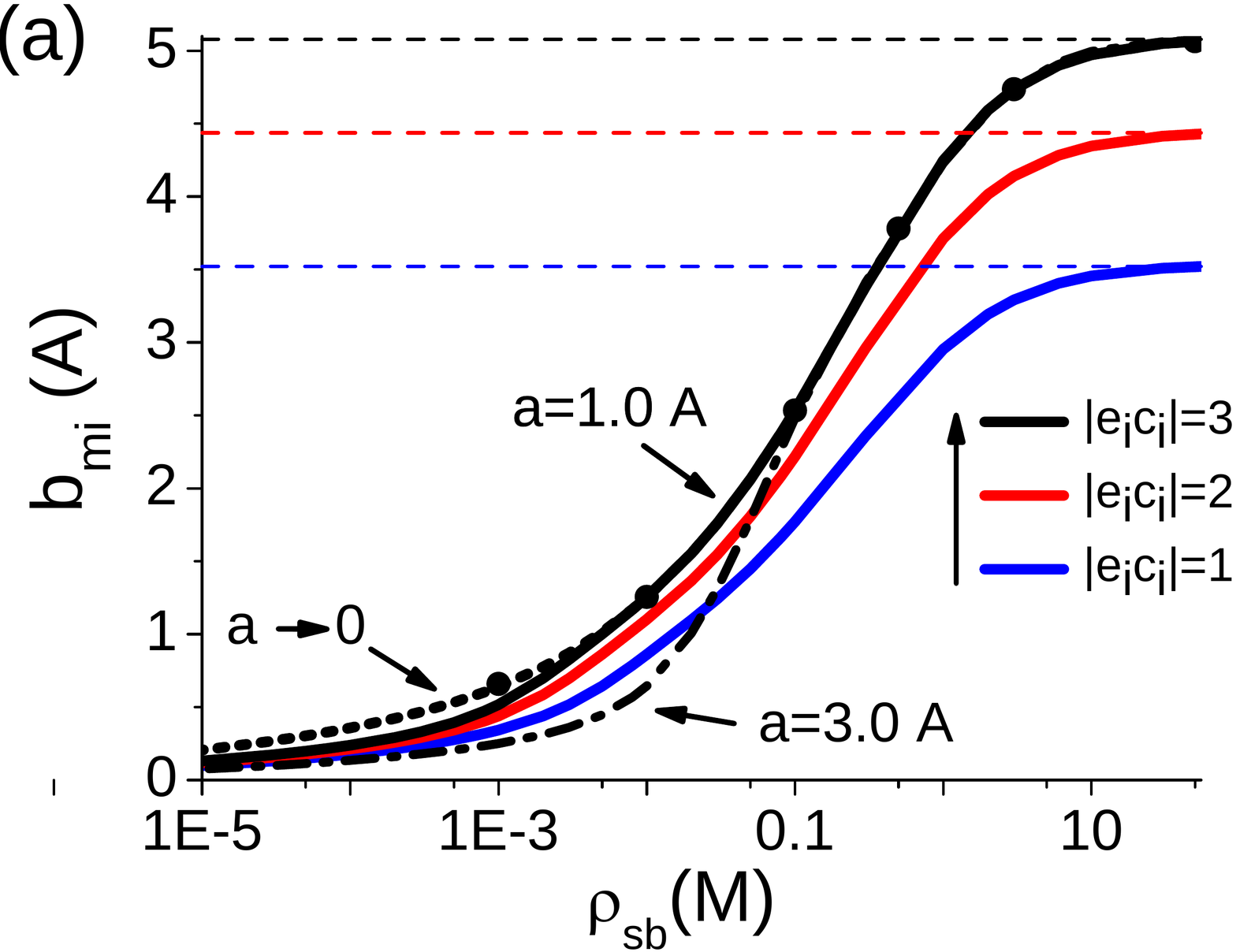}
\includegraphics[width=.47\linewidth]{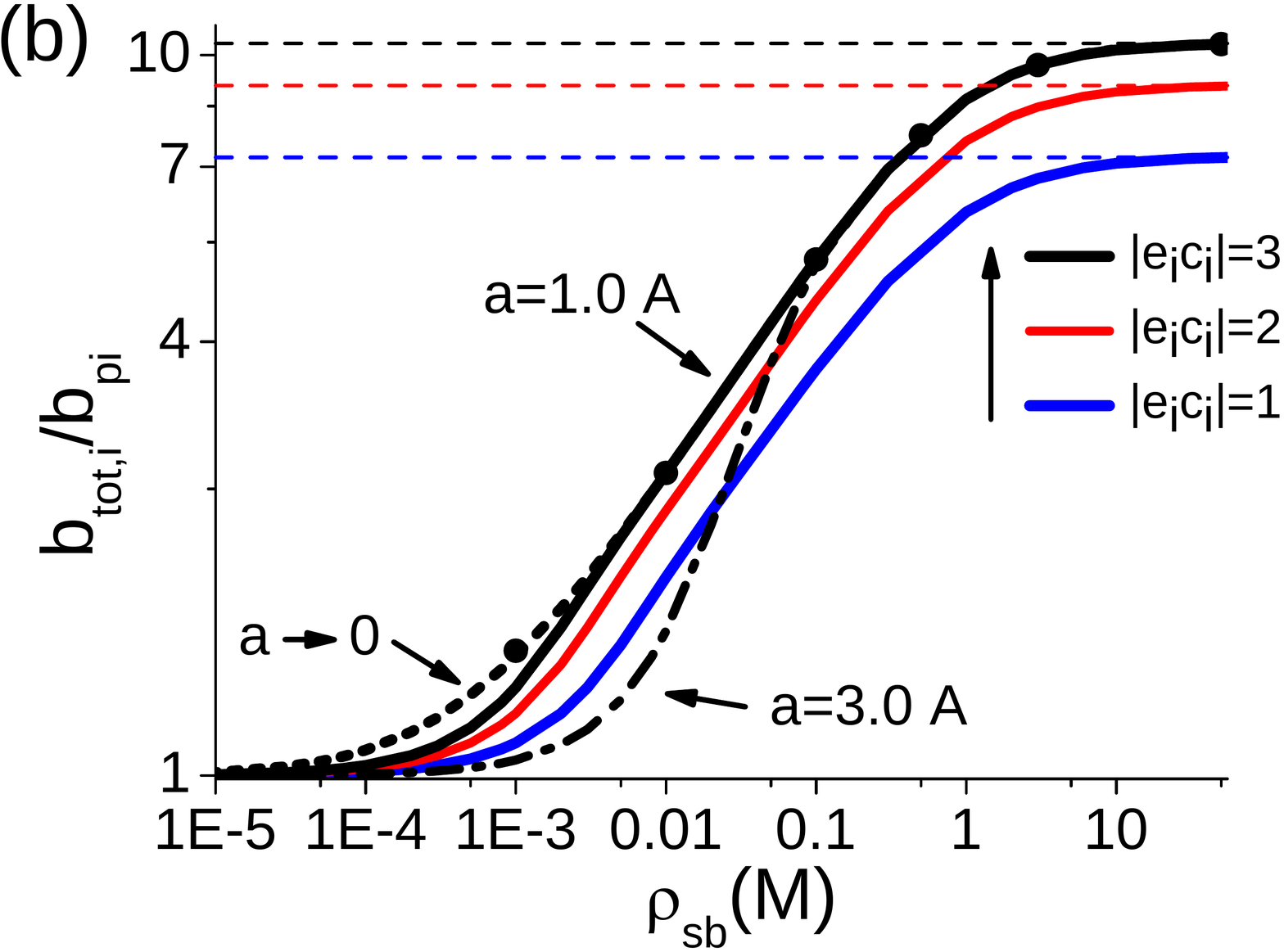}
\includegraphics[width=.47\linewidth]{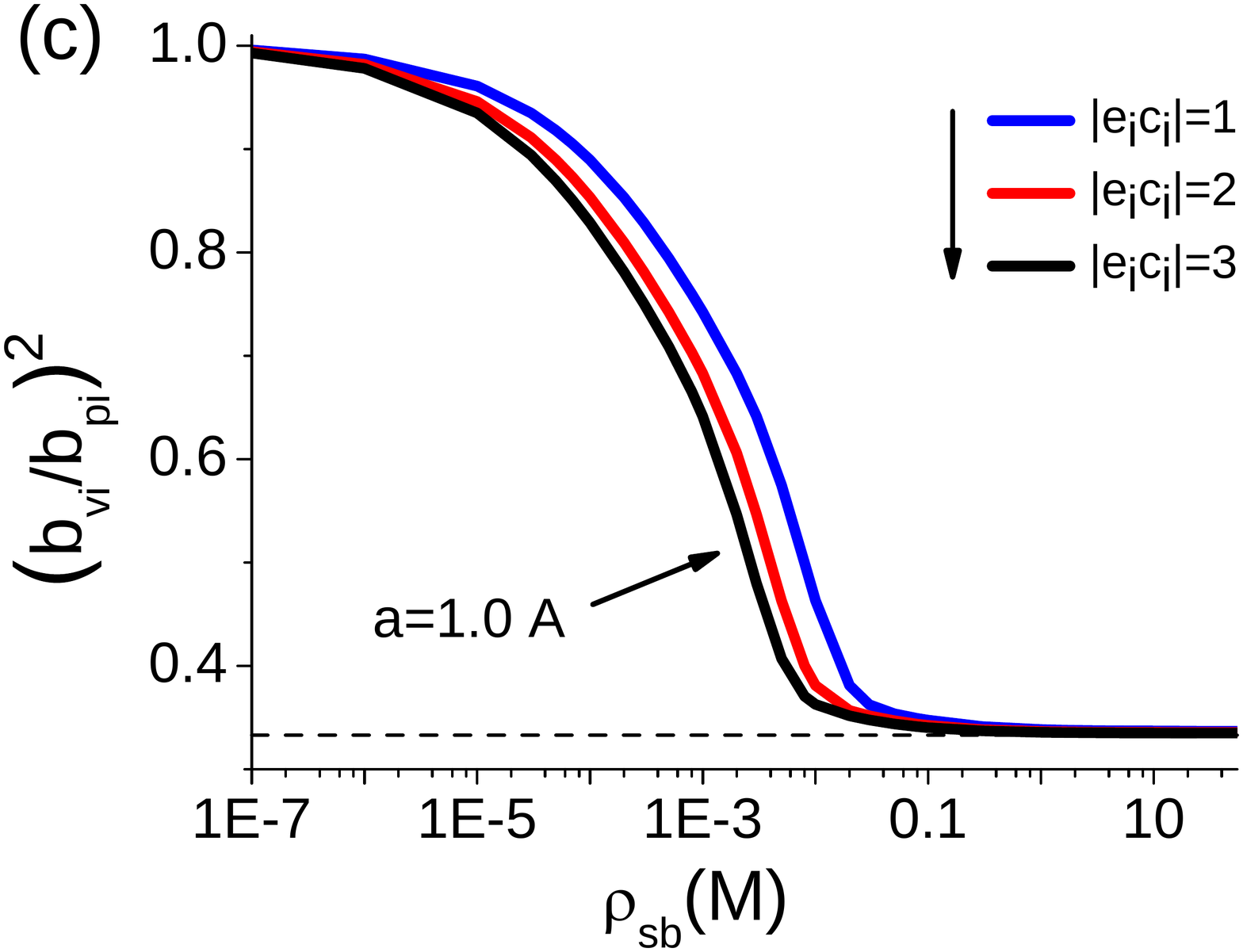}
\includegraphics[width=.47\linewidth]{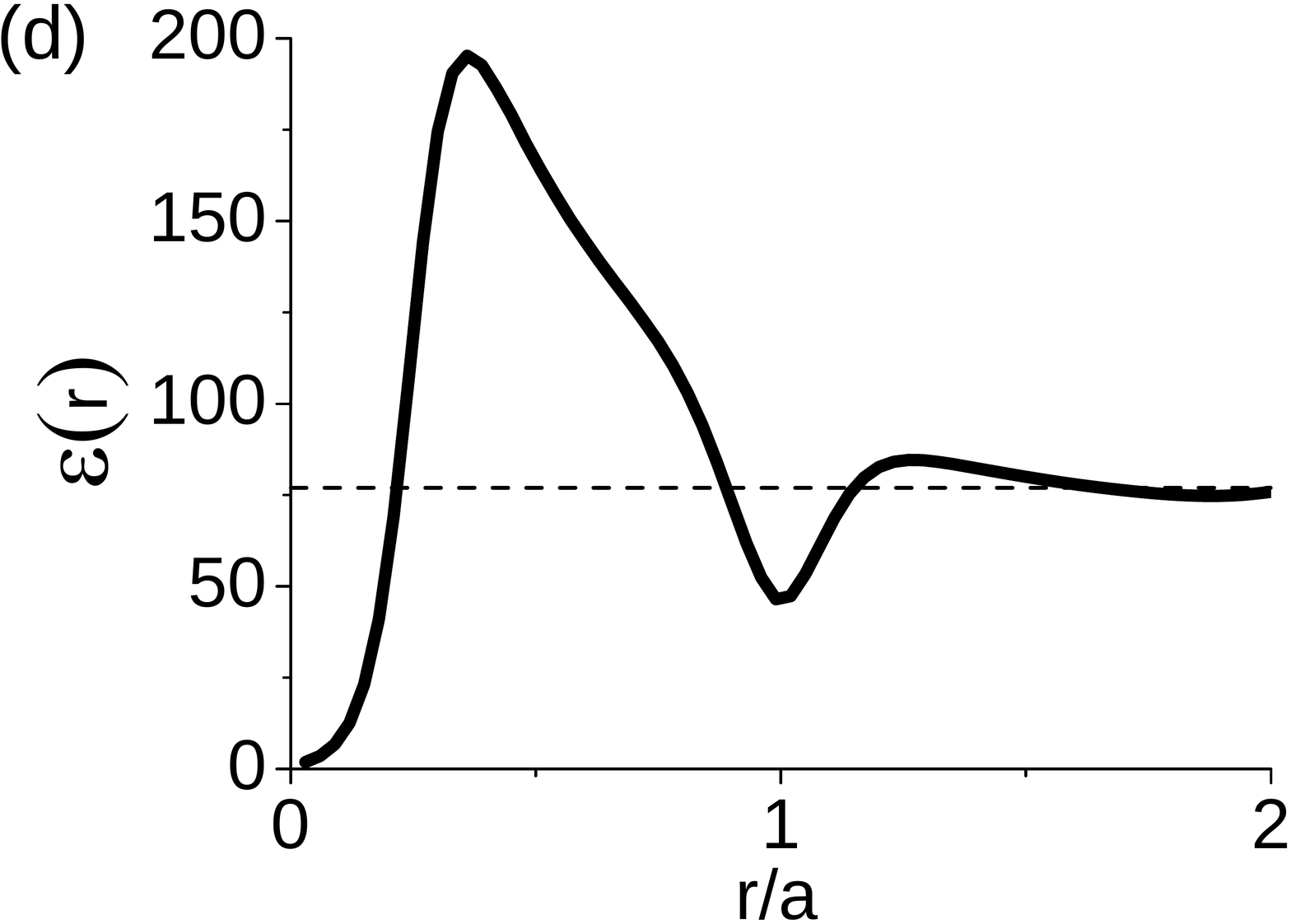}
\caption{(Color online) (a) Enhancement of the ionic dipole moment $b_{mi}$ introduced in Eq.~(\ref{bm1}) and (b) the total polarizability $b_{tot,i}$ defined in Eq.~(\ref{btot}), and (c) the reduction of the effective intrinsic polarizability $b_{vi}$ of Eq.~(\ref{bm2}) with increasing solvent density. The ion in the polar solvent has gas phase polarizability $b_{pi}=0.20$ {\AA}, and different ionic valencies from $|e_ic_i|=1$ to $3$ are considered. The results obtained from the numerical solution of the self-consistent equations~(\ref{varpot})-(\ref{dippmf}) at a fixed dipole moment $p_0=1$ {\AA} are displayed by solid curves for the solvent molecular size $a=1$ {\AA} and ion valencies $|e_ic_i|=1$ to $3$, and by dash-dotted black curves for $a=3$ {\AA} and divalent molecules with $|e_ic_i|=3$. Dotted curves in (a) and (b) denote for divalent molecules the point dipole results of Eq.~(\ref{ptd}) obtained in the limit $a\to0$ of Eqs.~(\ref{varpot})-(\ref{dippmf}) at fixed dipole moment, and circles mark the asymptotic equations~(\ref{as1}) and~(\ref{as2}) derived in the same point dipole limit for large solvent concentrations. Dashed horizontal curves correspond to the complete ionic hydration state of Eqs.~(\ref{bmsat})-(\ref{btotsat}). (d) Dielectric permittivity profile around a point ion at $r=0$ for the solvent density $\rho_{sb}=55$ M.}
\label{fig2}
\end{figure*}

To extend the investigation of the hydration induced modification of the electronic cloud radius and rigidity beyond the dilute solvent regime, we can map Eqs.~(\ref{varpot}) and~(\ref{dippmf}) onto an effective polarizable ion model. By adsorbing the effect of the hydration potential $\psi_{ib}(b)$ into an effective Drude oscillator model 
\be\label{disef}
h_i(\bb)=\frac{(\bb-\bb_{mi})^2}{4b^2_{vi}}, 
\ee
with the average dipole moment (or electronic cloud radius) $b_{mi}$ and induced ion polarizability $b_{vi}$ in the liquid environment, and evaluating the average in Eq.~(\ref{varep}) with the distortion potential~(\ref{disef}) without the hydration PMF~(\ref{dippmf}), we are left with the effective permittivity function
\bea\label{mf2}
\te_{eff}(q)&=&1+\frac{\kappa_s^2}{q^2}\left[1-\frac{\sin(qa)}{qa}\right]\\
&&\hspace{.3cm}+\sum_i\frac{\kappa_{ip}^2}{q^2}\left[1-\frac{\sin(qb_{mi})}{qb_{mi}}e^{-b^2_{vi}q^2}\right].\nonumber
\eea
The comparison of the function~(\ref{mf2}) with Eq.~(\ref{varepmf}) indicates that at the MF level, the ion has no dipole moment ($b_{mi}=0$), and its polarizability is equal to the gas phase value ($b_{vi}=b_{pi}$). By expanding now Eqs.~(\ref{varep}) and~(\ref{mf2}) in the infrared (IR) regime up to the order $O(q^4)$ and identifying the quadratic and quartic terms in the wavevector $q$, one obtains the coupled equations $6b^2_{vi}+b^2_{mi}=\lan b^2\ran$ and $60b^4_{vi}+20b^2_{mi}b^2_{vi}+b^4_{mi}=\lan b^4\ran$. The solution of these equations respectively yields for the average dipole moment  and induced polarizability of the hydrated ion
\bea\label{bm1}
&&b^2_{mi}=\left[\frac{5}{2}\lan b^2\ran^2-\frac{3}{2}\lan b^4\ran\right]^{1/2}\\
\label{bm2}
&&b^2_{vi}=b^2_{tot,i}-\frac{b^2_{mi}}{6},
\eea
where we introduced the total ionic polarizability
\be
\label{btot}
b^2_{tot,i}=\frac{1}{6}\lan b^2\ran.
\ee

We evaluated the dipole moment and polarizabilities in Eqs.~(\ref{bm1})-(\ref{btot}) with the numerical solution of Eqs.~(\ref{varpot})-(\ref{dippmf}). Figure~\ref{fig2}(a) displays the variation of the ionic dipole moment $b_{mi}$  with solvent density for the gas phase polarizability $b_{pi}=0.2$ {\AA} and various molecular valencies (solid curves). First of all, it is seen that an increase of the solvent concentration is accompanied with a monotonic rise of the dipole moment from zero to $b_{mi}\simeq 4$ {\AA}, until the latter saturates in the density regime $\rho_{sb}> 10$ M where the ion becomes fully hydrated. Then, one notices in Fig.~\ref{fig2}(b) that the expansion of the average electronic cloud radius upon hydration results in turn in an amplification of the total polarizability $b_{tot,i}$ by several factors. We note that the increase of the ionic polarizability upon hydration in a high dielectric liquid has been previously observed in ab-initio calculations with PCM solvent~\cite{ionpol,ionpol2}. This peculiarity was also revealed in Ref.~\cite{watwat} for water molecules, whose transfer from gas to liquid state was shown to be accompanied with a large amplification of their average dipole moment. In Section~\ref{ion}, it will be shown that a similar hydration mechanism is present as well in ionic liquids.

Moreover, in Fig.~\ref{fig2}(c), one sees that the effective intrinsic polarizability $b_{vi}$ exhibits in turn a monotonic decrease upon hydration, until it reaches in the fully hydrated state almost half of its gas phase value $b_{pi}$.  This indicates that upon hydration, the electronic cloud of the polarizable molecule increases in size, but also reaches an enhanced rigidity. In other words, the hydration opposes the electronic cloud deformation resulting from thermal fluctuations. Interestingly, comparison of Figs.~\ref{fig2} (a) and (c) shows that the increase of the electron cloud rigidity manifests itself at considerably lower concentrations than its expansion. Furthermore, in Figs.~\ref{fig2} (a) and (b), one notices that a significant departure from the MF behavior with $b_m=0$ and $b_{tot,i}=b_{pi}$ is observed above the characteristic solvent concentration $\rho_{sb}\simeq 10^{-3}$ M. This shows that in Fig.~\ref{fig2}(c), the hardening of the electronic cloud takes place already in the weak electrostatic coupling regime. Finally, in Figs.~\ref{fig2} (a)-(c), we note  that although ions with a higher valency are clearly better solvated, the ionic dipole moment and polarizabilities exhibit weaker sensitivity to the molecular charge than the hydration energy in Eq.~(\ref{dippmf}) characterized by a linear dependence on the charge $|e_ic_i|$.

\begin{figure}
\includegraphics[width=1.0\linewidth]{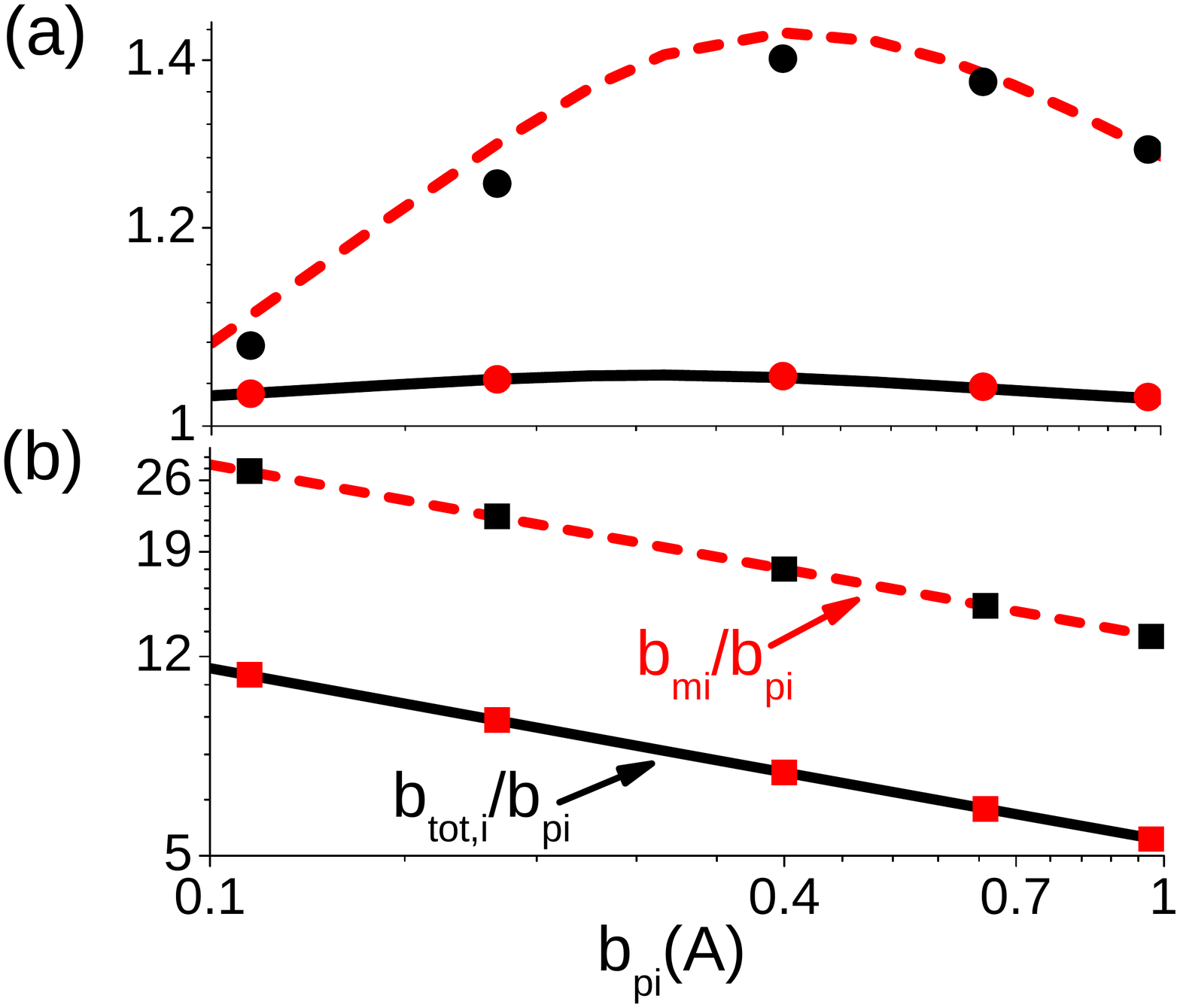}
\caption{(Color online) Rescaled ionic dipole moment (dashed red curves) and total polarizability (solid black curves) against the gas phase polarizability $b_{pi}$ for the solvent densities (a) $\rho_{sb}=2.0\times10^{-4}$ M and (b) $\rho_{sb}=55.0$ M, molecular charge $|e_ic_i|=2$, and solvent molecular size $a=1$ {\AA}. The curves are from the full numerical calculation, the black and red circles respectively correspond to the limiting laws~(\ref{bmdil}) and~(\ref{b2dil}), and the black and red squares are from the expressions~(\ref{bmsat}) and~(\ref{btotsat}) for the fully hydrated state.}
\label{fig3}
\end{figure}

In order to characterize the scaling of the hydrated polarizabilities with the gas phase polarizability $b_{pi}$, we first consider the electrostatic weak coupling regime of dilute solvents. By evaluating in the dilute solvent regime the averages in Eqs.~(\ref{bm1}) and~(\ref{btot}) at the order $O\left((\kappa_sa)^2\ell_B/a\right)$, one obtains for the ionic dipole moment and the total polarizability
\bea\label{bmdil}
\frac{b_{mi}^4}{b^4_{pi}}&=&\frac{12|e_ic_i|}{\sqrt\pi}\frac{\ell_B}{a}\left(\kappa_sa\right)^2f\left(\frac{a}{b_{pi}}\right)\\
\label{b2dil}
\frac{b_{tot,i}}{b_{pi}}&=&1+\frac{|e_ic_i|}{3\sqrt\pi}\frac{\ell_B}{a}\left(\kappa_sa\right)^2g\left(\frac{a}{b_{pi}}\right),
\eea
where we introduced the auxiliary functions
\bea
f(x)&=&x^{-1}\left[1-e^{-x^2/4}\right]\\
g(x)&=&x^{-1}-x^{-2}\sqrt\pi\;\mathrm{Erf}\left(\frac{x}{2}\right).
 \eea
In Fig.~\ref{fig3}(a), we compare the prediction of these asymptotic laws (circles) with the numerical solution of Eqs.~(\ref{varpot})-(\ref{dippmf}) (continuous curves) for a dilute liquid with density $\rho_{sb}=2.0\times10^{-4}$ M. One notices that the behavior of the polarizabilities is characterized by two regimes separated by a peak located at $b_{pi}\simeq a/3$. Indeed, the asymptotic limit of Eqs.~(\ref{bmdil}) and~(\ref{b2dil}) indicate that the average electronic cloud radius and total polarizability grow with the gas phase polarizability as $b_{mi}\sim b_{pi}^{5/4}$ and $b_{tot,i}-b_{pi}\sim b_{pi}^2$ for $b_{pi}\ll a/3$ (left branch of the curves in Fig.~\ref{fig3}(a)), and $b_{mi}\sim b_{pi}^{3/4}$ and $b_{tot,i}-b_{pi}\sim c^{st}$ for $b_{pi}\gg a/3$ (right branch of the curves). Thus, the transition between these two regimes results from a competition between the solvent molecular size and the gas phase polarizability.

In the opposite regime of concentrated solvents, the expansion of Eqs.~(\ref{varpot}) and~(\ref{dippmf}) for $\kappa_sa\gg1$ and $b_p/a\ll1$ yields for the hydration energy the asymptotic limit
\be
\psi_{ip}(b)\simeq-\frac{|e_ic_i|\ell_B}{b}\left[e^{-\kappa_sb}+\kappa_sb-1\right].
\ee
Neglecting the exponential term and expanding the total distortion potential $U_i(b)=h_{i}(b)+\psi_{ip}(b)$ around the equilibrium position, we are left with the gaussian distribution $U_i(b)=\left(b-b_{mi}\right)^2/\left(4b_{vi}^2\right)$, with the average electronic cloud radius and effective intrinsic polarizability
\bea\label{bmsat}
\frac{b_{mi}}{b_{pi}}&=&\left(\frac{2|e_ic_i|\ell_B}{b_{pi}}\right)^{1/3}\\
\label{bvsat}
\frac{b_{vi}^2}{b_{pi}^2}&=&\frac{1}{3}.
\eea
Substituting these relations into Eq.~(\ref{bm2}), the total ionic polarizability follows as
\be\label{btotsat}
\frac{b_{tot,i}^2}{b_{pi}^2}=\frac{1}{3}\left[1+\frac{1}{2}\left(2|e_ic_i|\frac{\ell_B}{b_{pi}}\right)^{2/3}\right].
\ee
Figures~\ref{fig2}(a)-(c) show that the closed form expressions in Eqs.~(\ref{bmsat})-(\ref{btotsat}) accurately reproduce the saturation values of the ionic dipole moment and the polarizabilities (dashed horizontal curves). First of all, in Eq.~(\ref{bvsat}), one notes that regardless of the ion charge, transferring  the ion from the gaseous phase into the liquid environment reduces its intrinsic polarizability by a factor three. Moreover, Eqs.~(\ref{bmsat}) and~(\ref{btotsat}) indicate that in the fully hydrated state, the ionic dipole moments and total polarizability grow as the cubic root of the ion charge, which explains the weak dependence of the solvation on the molecular charge strength in Figs.~\ref{fig2}(a)-(c).

We compare in Fig.~\ref{fig3}(b) the limiting laws~(\ref{bmsat}) and~(\ref{btotsat}) with the full numerical solution of the self-consistent equations for the solvent concentration $\rho_{sb}=55.0$ M. These equations indicate that in the range $b_{pi}=0.1$ {\AA} to $1.0$ {\AA}, the dipole moment and polarizability of the fully hydrated ion grows with the gas phase polarizability according to the $b_{pi}^{2/3}$ power law. We also note that interestingly, the hydrated polarizabilities in Eqs.~(\ref{bmsat})-(\ref{btotsat}) are independent of the solvent molecular size. This peculiarity stems from the fact that the complete hydration takes place in the parameter regime $\kappa_s\gg a^{-1}$, where the part of the dielectric susceptibility function associated with the rotation of solvent molecules (i.e. the third term on the r.h.s. of Eq.~(\ref{varep})) makes no contribution to the hydration energy $\psi_{ib}(b)$ in Eq.~(\ref{dippmf}).

In our previous work on the MF theory of polar liquids at charged interfaces, it was shown that the non-local character of electrostatic interactions in the solvent results from the finite size of solvent molecules~\cite{nlpb}. The effect of non-locality on the hydration mechanism can be estimated by varying the solvent molecular size $a$ at fixed dipole moment $p_0=Qa$. To this aim, we reexpress the dielectric permittivity function~(\ref{varep}) in the form
\be\label{perres}
\te(q)=1+\frac{\left(\kappa_sp_0\right)^2}{\left(Qqa\right)^2}\left[1-\frac{\sin(qa)}{qa}\right],
\ee
and calculate the total polarizabilities~(\ref{bm1})-(\ref{btot}) with the above permittivity function by varying $a$ with the dipole moment fixed at $p_0=1$ {\AA}. In Figs.~\ref{fig2} (a) and (b), the comparison of the curves with $a=1$ {\AA} and $3$ {\AA}  shows that the increase of the solvent molecular size at fixed dipole moment lowers the average electronic cloud radius and the total ionic polarizability. Hence, non-locality weakens the hydration of the polarizable ion. To explain this peculiarity, we note that in the dilute ion regime, the inverse Fourier transform of the potential in Eq.~(\ref{varpot}) is given by a generalized Coulomb law, $v_0(r)=\ell_B/\left[r\e(r)\right]$, with the local dielectric permittivity function
\be\label{locdi}
\e(r)=\frac{\pi}{2}\left/\int_0^\infty\frac{\mathrm{d}k}{k}\frac{\sin(kr/a)}{\te(k)}\right.,
\ee
and the adimensional wavevector $k=qa$. The dielectric permittivity profile of Eq.~(\ref{locdi}) is reported in Fig.~\ref{fig2}(d). First of all, it is seen that the close vicinity of the ion at $r=0$ is characterized by a dielectric void. Then, one notes that the dielectric permittivity function in Eq.~(\ref{locdi}) depends solely on the rescaled distance $r/a$. This means that an increase of the solvent molecular size amplifies the dielectric void around a polarizable molecule, and consequently reduces its hydration energy in Eq.~(\ref{dippmf}).

In the opposite point-dipole limit of solvent molecules $a\to 0$, the permittivity function~(\ref{perres}) tends to the bulk permittivity, $\te(q)\to\e_b=1+4\pi\ell_Bp_0^2\rho_{sb}/3$, and the hydration PMF~(\ref{dippmf}) takes the simple form~\cite{rem2}
\be\label{psipt}
\psi_{ip}(b)=\psi_{ip}(b\to\infty)+\frac{4\Gamma b_{pi}}{b},
\ee
with the adimensional parameter
\be\label{coup}
\Gamma=\frac{\left(\kappa_sa\right)^2}{6+\left(\kappa_sa\right)^2}\frac{|e_ic_i|\ell_B}{4b_{pi}}.
\ee
Evaluating the integrals in Eq.~(\ref{av}) with the PMF~(\ref{psipt}), the moments of the electronic cloud oscillations can be expressed in terms of Meijer G-functions~\cite{math},
\be\label{ptd}
\frac{\lan b^{2n}\ran}{b_{pi}^{2n}}=\left(2\Gamma\right)^{2n}\frac{\mathrm{G}_{03}^{30}\left(-n-\frac{3}{2},-n-1,0\left|\Gamma^2\right.\right)}{\mathrm{G}_{03}^{30}\left(-\frac{3}{2},-1,0\left|\Gamma^2\right.\right)}.
\ee
The ionic dipole moment and total polarizability obtained from Eq.~(\ref{ptd}) is reported in Figs.~\ref{fig2}(a) and (b). One notices that the point dipole result is very close to the case with finite solvent molecular size $a=1$ {\AA}. Thus, for the model parameters chosen in this work, non-locality plays a minor role in the hydration process. It is interesting to note that in this parameter regime, the hydration of the polarizable ion can be solely described by the single coupling parameter $\Gamma$.

By Taylor-expanding Eq.~(\ref{ptd}) in the regime $\Gamma\gg1$, one obtains for the ionic dipole moment and total polarizability the following expressions,
\bea
\label{as1}
&&\frac{b_{mi}}{b_{pi}}=2\Gamma^{1/3}+\frac{2}{3}\Gamma^{-1/3}+O\left(\Gamma^{-1}\right)\\
\label{as2}
&&\frac{b_{tot,i}}{b_{pi}}=\sqrt{\frac{2}{3}}\Gamma^{1/3}+\frac{7}{6\sqrt{6}}\Gamma^{-1/3}+O\left(\Gamma^{-1}\right).
\eea
In Figs.~\ref{fig2}(a) and (b), it is shown that the asymptotic laws~(\ref{as1}) and~(\ref{as2}) can accurately reproduce the increase of the ionic dipole moment and total polarizability from $\rho_{sb}=10^{-3}$ M to complete hydration.  These equations indicate that the fully hydrated state of the polarizable ion is reached with increasing solvent concentration through the gradual saturation of the parameter $\Gamma$ in Eq.~(\ref{coup}). We consider next the counterpart of this hydration process in ionic liquids without solvent molecules.

\subsection{Cooperative solvation in ionic liquids}
\label{ion}
\begin{figure}
\includegraphics[width=1.0\linewidth]{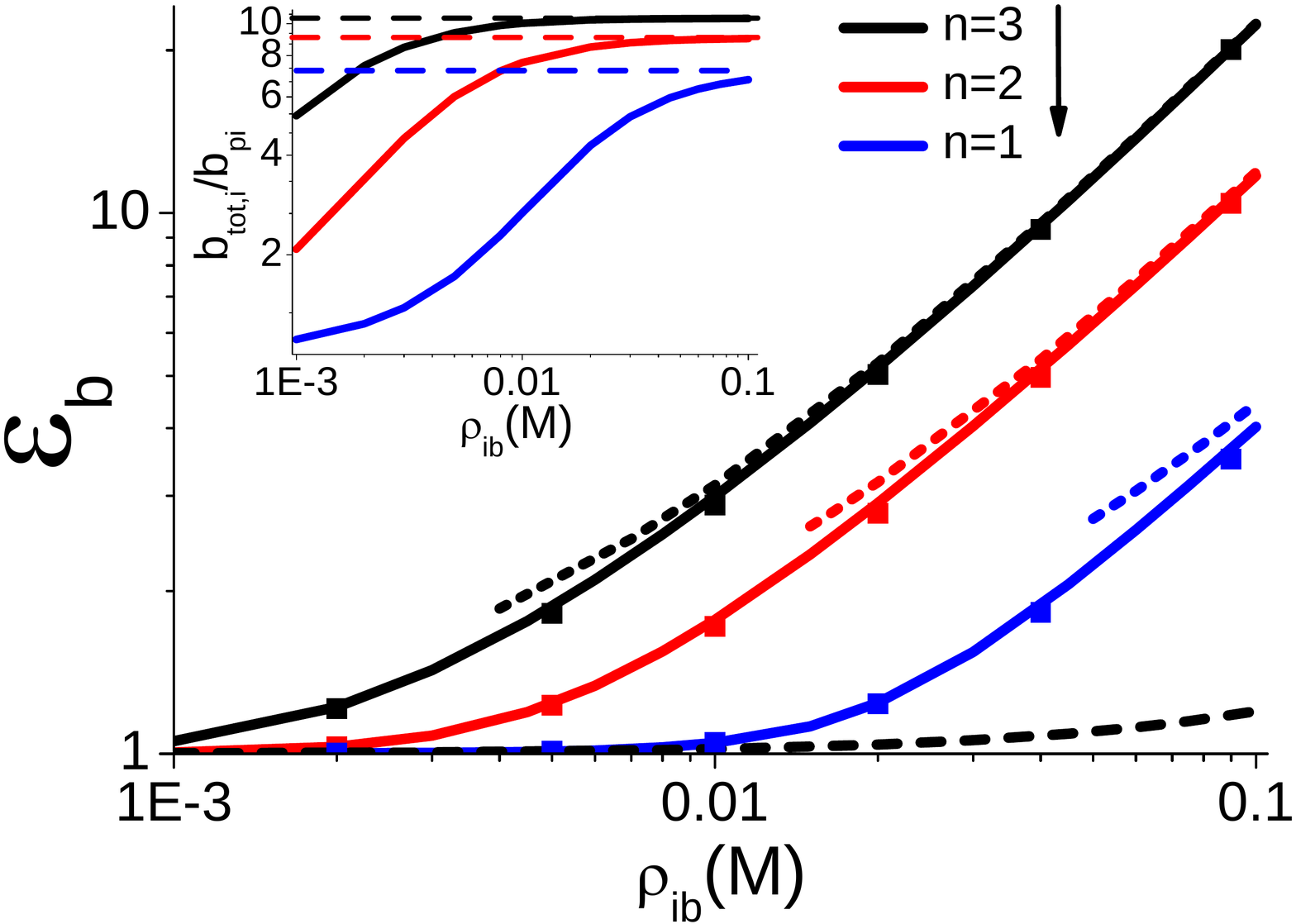}
\caption{(Color online) Effective dielectric permittivity of ionic liquids with bare polarizability $b_{pi}=0.2$ {\AA} and valency $n$. Solid curves are from the numerical solution of Eqs.~(\ref{varpot})-(\ref{dippmf}), dotted curves denote the full solvation limit in Eq.~(\ref{dielsat}), and square symbols from the approximative scheme of Eqs.~(\ref{ptpol1})-(\ref{ptpol2}). The black dashed curve is the MF dielectric permittivity $\e_b=1+\xi_p$ for $n=3$. Inset : Total ionic polarizabilities (solid curves) and their saturation values from Eq.~(\ref{btotsat}) (dashed horizontal curves).}
\label{fig4}
\end{figure}

Ionic liquids are promising salt solvents that gradually replace water in new generation energy storage devices such as graphene based capacitors~\cite{gr}. The accurate knowledge of the dielectric permittivity of ionic liquids is needed to predict the charge storage ability of these devices. In ab-initio calculations of ionic liquids composed of small ions with negligible dipole moments~\cite{ionion1}, it was found that the contribution from electronic and orientational polarization of individual ions cannot alone explain the large dielectric permittivities measured in experiments~\cite{crc}. Based on this observation, it was also argued that an additional polarization effect induced by the surrounding ions must be present to explain the high dielectric permittivity values.

In order to shed light on this point, we consider in this part the closure equations~(\ref{varpot})-(\ref{dippmf}) for an ionic liquid free of solvent molecules, and composed of two species of polarizable ions with the same bare polarizability $b_{pi}$ and bulk density $\rho_{ib}$. Furthermore, the point charges on the polarizable molecules are $e_{1,2}=\pm1$ and $c_{12}=\pm n$, which corresponds to the net molecular charges $q_{1,2}=\pm(n-1)$  (see Fig.~\ref{fig1}(b)). The dielectric permittivity of the medium at large separation distances from a central ion is obtained from the IR limit of Eq.~(\ref{varep}), $\e_b=\te(q\to0)$, and it is given by
\be\label{diel}
\e_b=1+\sum_i\kappa_{ip}^2b_{tot,i}^2,
\ee
where the total ionic polarizability defined in Eq.~(\ref{btot}) has to be computed from the numerical solution of Eqs.~(\ref{varpot})-(\ref{dippmf}). Indeed, for an ionic liquid where the hydration of the polarizable ion affects the polarization of the surrounding medium in a self-consistent way, the solution of these equations is more tricky. Our numerical scheme consisted in solving these equations by iteration on a discretized Fourier lattice. Namely, at the first iterative level, the MF permittivity of Eq.~(\ref{varepmf}) was used as the input function in the potential Eq.~(\ref{varpot}) in order to evaluate the hydration PMF in Eq.~(\ref{dippmf}), and the latter was injected at the next step into Eq.~(\ref{av}) to obtain the updated dielectric permittivity function from Eq.~(\ref{varep}). This procedure was continued until self-consistency was achieved.

We illustrate in Fig.~\ref{fig4} the ionic polarizability (inset) and the dielectric permittivity of the liquid (main plot) obtained from the numerical solution of Eqs.~(\ref{varpot})-(\ref{dippmf}) (solid curves). First of all, it is seen that the increase of the ion density is accompanied with a strong amplification of the total ion polarizability, which in turn results in a rise of the dielectric permittivity of the medium. Then, in the inset of Fig.~\ref{fig4}, we note that unlike the case of a polarizable ion in a polar solvent (see Fig.~\ref{fig2}(a)), the ionic polarizability and the full hydration density exhibits a pronounced dependence on the molecular charge.

\begin{figure}
\includegraphics[width=1.0\linewidth]{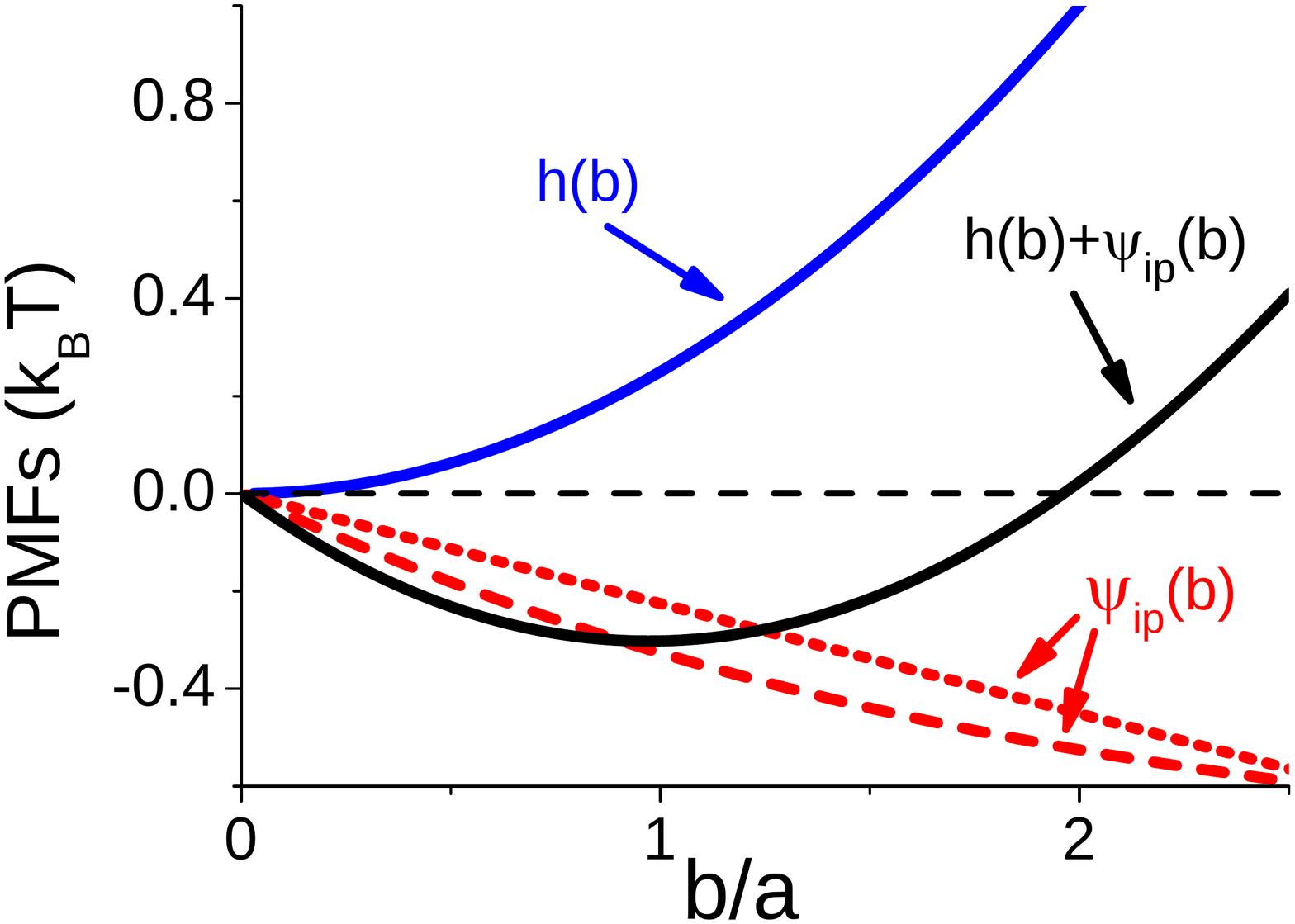}
\caption{(Color online) Drude oscillator potential Eq.~(\ref{hpol}) (blue curve), the first (charge screening) and second term (dielectric screening) of the ionic solvation energy in Eq.~(\ref{ionsolv}) denoted respectively by the red dotted and dashed curves, and the total distortion potential (black curve) for an ionic liquid composed of polarizable ions only, with ionic density per species $\rho_{ib}=5\times10^{-5}$ M, gas phase polarizability $b_{pi}=1$ {\AA}, and molecular charge $|e_ic_i|=2$.}
\label{fig5}
\end{figure}

These effects can be shown to be driven by the self-consistent solvation of polarizable ions by their own field. To this aim, we introduce respectively the charge and dipolar screening parameters
\bea
\kappa_c^2&=&4\pi\ell_B\sum_i\rho_{ib}q_i^2=8\pi\ell_B(n-1)^2\rho_{ib}\\
\kappa_p^2&=&\sum_i\kappa_{ip}^2=16\pi\ell_Bn\rho_{ib},
\eea
and the corresponding coupling parameters $\xi_c=\left(\kappa_cb_{pi}\right)^2$ and $\xi_p=\left(\kappa_pb_{pi}\right)^2$. In the dilute liquid regime, by expanding the closure relations~(\ref{varpot}) and~(\ref{dippmf}) up to the order $O\left(\xi_c\right)$ and $O\left(\xi_p\right)$, one obtains for the solvation PMF
\be\label{ionsolv}
\psi_{ip}(b)=-\frac{|e_ic_i|\ell_B}{2b_{pi}}\left[\xi_c\frac{b}{b_{pi}}+\xi_pF\left(\frac{b}{b_{pi}}\right)\right],
\ee
where we introduced the auxiliary function
\be
F(x)=1+\sqrt\pi\frac{x}{4}-\frac{1}{2}e^{-x^2/4}-\sqrt\pi\frac{x^2+2}{4x}\mathrm{Erf}\left(\frac{x}{2}\right).
\ee
One sees in Eq.~(\ref{ionsolv}) that the solvation energy is composed of a contribution from the charge screening (the first term on the r.h.s.), and a part resulting from the polarizability induced dielectric screening of the ion by the surrounding ionic liquid (the second term on the r.h.s.). We display in Fig.~\ref{fig5} the PMF of Eq.~(\ref{ionsolv}) for  a monovalent ionic solution ($n=2$) with concentration $\rho_{ib}=5\times10^{-5}$. It is seen that in this dilute liquid regime, the charge and dielectric screening effects independently lower the bare distortion energy $h_i(b)$ with an equal weight, thus favoring the expansion of the electronic cloud. Then, we note that as in the case of a polarizable ion in a polar solvent considered in Section~\ref{polar}, the total distortion potential exhibits a minimum. In other words, in the liquid environment, the polarizable ion acquires a finite dipole moment. We emphasize that this effect has been previously observed in ab-initio calculations of ionic liquids composed of charges with fluctuating geometry~\cite{ionion1}.

In order to determine the relative weight of the dielectric and charge screening mechanisms in the renormalization of the background dielectric permittivity beyond the dilute regime, we will introduce an approximative solution scheme of Eqs.~(\ref{varpot})-(\ref{dippmf}). To this aim, we first redefine the hydration PMF of Eq.~(\ref{dippmf}) by subtracting the constant energy in the dissociated state, $\varphi_{ip}(b)=\psi_{ip}(b)-\psi_{ip}(b\to\infty)$. Introducing the dimensioneless wavevector $k=b_{pi}q$ and separation distance $x=b/b_{pi}$, this PMF can be expressed as
\bea
\label{dippmf2}
\varphi_{ip}(x)&=&\frac{|e_ic_i|\ell_B}{b_{pi}}\frac{2}{\pi}\int_0^\infty\mathrm{d}k
\frac{\xi_p\lan1-\frac{sin(kx')}{kx'}\ran\frac{sin(kx)}{kx}}{k^2+\xi_c+\xi_p\lan1-\frac{sin(kx')}{kx'}\ran},\nonumber\\
\eea
where the statistical average of the functions inside the brackets is still evaluated according to Eq.~(\ref{av}) with the adimensional electronic cloud radius $x'=b'/b_{pi}$ as the integration variable. We now assume that the hydration PMF affects the electron cloud mainly at small separations $x<1$. This implies that in Eq.~(\ref{dippmf2}), only small wavevectors $k<1$ make a significant contribution to the integral. Based on this assumption, by expanding the sinusodidal functions inside the bracket of Eq.~(\ref{dippmf2}) at the order $O\left(k^2\right)$, the integral can be evaluated exactly. Within this approximation, the complicated integral equations~(\ref{varpot})-(\ref{dippmf}) for the dielectric permittivity are reduced to a simpler non-linear equation,
\bea\label{ptpol1}
&&\e_b=1+\frac{\xi_p}{6}\frac{\int_0^\infty\mathrm{d}xx^4e^{-x^2/4-\varphi_{ip}(x)}}{\int_0^\infty\mathrm{d}xx^2e^{-x^2/4-\varphi_{ip}(x)}}\\
\label{ptpol2}
&&\varphi_{ip}(x)=\frac{\e_b-e^{-x\sqrt{\xi_c/\e_b}}}{\e_b}\frac{|e_ic_i|\ell_B}{b_{pi}x},
\eea
where we made use of Eqs.~(\ref{av}) and~(\ref{diel}).

In Fig.~\ref{fig4}, it is shown that the numerical solution of Eq.~(\ref{ptpol1}) can accurately reproduce the dielectric permittivity obtained from the closure equations~(\ref{varpot})-(\ref{dippmf}) over the whole density range. We now note that in the solvation PMF of Eq.~(\ref{ptpol2}), the contribution from the dielectric and charge screenings correspond respectively to the first constant term $\e_b$ and the second exponential function in the numerator. This equation indicates that while increasing the ion concentration from the dilute regime, the exponential term is gradually dominated by the constant term in the numerator and becomes negligible for $\e_b\gg 1$. Thus, charge screening makes a significant contribution to the dielectric permittivity exclusively at low ion concentrations.

To asccertain the latter point, we now consider the strict limit of large liquid densities with $\kappa_{p}b_{pi}\gg1$. By evaluating the PMF of Eq.~(\ref{dippmf}) in this limit, we found that the total ionic polarizability is still given by the expression~(\ref{btotsat}) (see the horizontal lines in the inset of Fig.~\ref{fig4}). Substituting this relation into Eq.~(\ref{diel}), one obtains the dielectric permittivity of the ionic liquid at the fully solvated state
\be\label{dielsat}
\e_b=1+\frac{\xi_p}{3}\left[1+\frac{1}{2}\left(2|e_ic_i|\frac{\ell_B}{b_{pi}}\right)^{2/3}\right].
\ee
In the main plot of Fig.~\ref{fig4}, it is shown that this closed form expression is a very good approximation for the dielectric permittivity of the ionic liquid beyond the dilute regime. One can note that in Eq.~(\ref{dielsat}), the dependence of the permittivity on the charge screening parameter $\xi_c$ has disappeared. This shows that close to the full solvation state, the collective solvation mechanism is solely driven by the dielectric screening induced by polarizable ions.

We also compare in Fig.~\ref{fig4} the MF level bulk dielectric permittivity $\e_b=1+\xi_p$ for the ion valency $n=3$ with the self-consistent result. The MF theory that neglects the collective ionic solvation is shown to strongly underestimate the dielectric permittivity of the ionic liquid. This observation is in line with Ref.~\cite{ionion2} where the rotational polarizability associated with the gas phase dipole moment of ions was shown to be unsufficient to explain the high dielectric permittivity of ionic liquids. This suggests that the cooperative hydration mechanism scrutinized in this part brings the main contribution to the dielectric permittivities of ionic liquids. Hence, correlation effects cannot be neglected in polarizable liquids.

\section{Conclusion}

We have introduced in this article a classical electrostatic theory of polarizable ions in high dielectric liquids. Within this theoretical framework, we have scrutinized the physical mechanism behind the ionic solvation properties observed in ab-initio calculations of polar solvents~\cite{ionpol,ionpol2} and ionic liquids~\cite{ionion1,ionion2}. In the first part of the article, we presented the electrostatic formulation of polarizable ions immersed in polar solvents composed of dipolar molecules with finite size. Then, we derived from the Dyson equation the electrostatic self-consistent relations accounting for the electrostatic correlations between the particles in the liquid.

The second part of the article was devoted to the hydration of a single polarizable in a polar solvent such as water. It was shown that the electrostatic energy release experienced by the polarizable ion upon hydration results in the expansion of its electronic cloud. As a result, the ion carrying zero dipole moment in the gas phase acquires in the liquid environment an average dipole moment. However, the hydration also amplifies the rigidity of the electronic cloud, thereby opposing its deformation induced by thermal fluctuations. In qualitative agreement with quantum molecular calculations with PCM solvent~\cite{ionpol,ionpol2}, the overall effect was shown to be an enhancement of the gas phase polarizability upon hydration.

In the third part of the article, we have investigated a cooperative solvation mechanism in ionic liquids free of solvent molecules. We have found that similar to the case of a polarizable ion in the polar solvent and in agreement with ab-initio calculations of ionic liquids~\cite{ionion1}, each polarizable ion acquires in the liquid a finite dipole moment and an increased polarizability. This effect resulting from the polarization field generated by the surrounding ions self-consistently amplifies the dielectric permittivity of the medium. We note that this solvation induced amplification of the dielectric permittivity is substantial even in the weak electrostatic coupling regime of dilute liquids. This suggests that the self-consistent solvation mechanism brings the dominant contribution to the dielectric permittivity of ionic liquids composed of small ions with negligible permanent dipole moment in the gas phase~\cite{ionion2}.

We have introduced the first microscopic theory of ionic hydration in explicit solvent, and we emphasize that the model as well as the theoretical scheme need refinements.  First of all, it should be noted that our approach does not account for the hydrogen bond formation in water solvent, which is believed to amplify the dielectric permittivity of water~\cite{ons}. This complication expected to become significant beyond the dilute liquid regime should be addressed in a future work by extending our approach beyond the gaussian field approximation, i.e. by opting for a more sophisticated closure to solve Eq.~(\ref{corr2}). An additional complication for solvents at physiological concentrations comes from the importance of excluded volume effects associated with the finite size of the particles in the liquid. The first step to generalize the model in this direction consists in including simple hard-core or repulsive Yukawa interactions between the particles as in Refs.~\cite{duny,jstat,jcp1}. Then, our theoretical scheme should be extended to a second order cumulant expansion of the grand potential around the reference Hamiltonian Eq.~(\ref{gauss}). This generalization would allow to determine how much our results are quantitatively modified beyond the dilute liquid regime. Indeed, we expect hard-core interactions between solvent molecules and ions to reduce the polarizability increase induced by the electrostatic hydration mechanism. In this sense, the results presented in this article beyond the dilute solvent regime should be considered as an upper boundary for the actual ionic cloud expansion effect. Our results should be also compared at the next step with MC simulations of the polarizable ion model introduced in Sec.~\ref{mod}, but these simulations are currently unavailable.

Finally, the consideration of the induced polarizability with a classical Drude potential is another limitation of the present model. Actually, it should be noted that the ionic dipole moments in the solvated state provided by our theory are larger than the values observed in ab-initio calculations~\cite{ionpol,ionpol2,ionion1}. For example, the Pauli exclusion effect neglected by the classical approach is expected to partially suppress the hydration induced expansion of the electron cloud. However, refinements at the quantum level are of course beyond the scope and the main message of the present work. Indeed, the ability of the theory to qualitatively capture ionic hydration effects observed in quantum molecular calculations for both polar solvents and ionic liquids on the one hand, and the presence of these effects in the dilute liquid regime where the complications discussed so far are not expected on the other hand confirm the physical consistency of the model with real dielectric liquids.
\\
\acknowledgements  This work has been in part supported by The Academy of Finland through its Centres of Excellence Program (project no. 251748) and NanoFluid grants.


\begin{thebibliography}{99}
\bibitem {rem} Our approximation in deriving these closure equations is equivalent to a variational minimization of the Grand potential with the reference Hamiltonian~(\ref{gauss}). That being said, starting with the Dyson equation presents itself as a shortcut, and the formal relation~(\ref{corr2}) for the two-point correlation function can be used in a future work to extend our approach beyond the gaussian field approximation.
\bibitem {rem2} We note that in the point-dipole limit $a\to 0$, the constant term $\psi_{ip}(b\to\infty)$ in Eq.~(\ref{psipt}) is divergent. Although this term could be regularized with an ultraviolet cut-off, this complication is unnecessary since this term does not affect the averages in Eq.~(\ref{av}).
\end{thebibliography}
\end{document}